\documentclass[twocolumn]{emulateapj}
\pdfoutput=1
\usepackage{multirow}
\usepackage{graphics}
\usepackage{amsmath, amsthm, amssymb}
\usepackage{epsfig}
\usepackage[hidelinks]{hyperref}
\usepackage{natbib}

\begin{document}

\title{A Simple formula for the third integral of motion\\ of disk-crossing stars in the Galaxy}

\author{Ronaldo S. S. Vieira$^{1,2}$}\email[e-mail: ]{ronssv@ifi.unicamp.br}

\affiliation{$^1$ Instituto de F\'{i}sica Gleb Wataghin, Universidade Estadual de
Campinas, 13083-859, Campinas, SP, Brazil}

\affiliation{$^2$ Nicolaus Copernicus Astronomical Center, ul. Bartycka, 18, 00-716 Warszawa, Poland}

\author{Javier Ramos-Caro$^3$}\email[e-mail: ]{javier@ufscar.br}

\affiliation{$^3$ Departamento de F\'{i}sica, Universidade Federal de S\~{a}o Carlos,
13565-905, S\~{a}o Carlos, SP, Brazil}


\begin{abstract}

We present an analytical simple formula for an approximated third integral of motion associated
with nearly equatorial orbits in the Galaxy:
$I_{3}=Z\Sigma_{I}^{1/3}$, where $Z(R)$ is the vertical amplitude of the orbit
at galactocentric distance $R$ and $\Sigma_{I}(R)$ is the integrated
dynamical surface mass density of the disk, a quantity which has recently become measurable.
We also suggest that this  relation is valid for disk-crossing orbits in a wide variety of axially symmetric
galactic models, which range from razor-thin disks to disks with non-negligible thickness,
whether or not the system includes bulges and halos.
We apply our formalism to a Miyamoto-Nagai model
and to a realistic model for the Milky Way. In both cases, the results provide fits for the shape
of nearly equatorial orbits which are better than the corresponding curves obtained by the usual adiabatic approximation when
the orbits have vertical amplitudes comparable to the disk's scale height. We also discuss the role of this
approximate third integral of motion in modified theories of gravity.
\end{abstract}

\maketitle

\section{Introduction}

The issue of the ``third integral of motion'' has captured the attention of the astrophysics community for decades.
We know nowadays the general form of an axially symmetric separable potential (see \citealp{dezeeuw88} and references therein)
and the vast majority of axially symmetric galactic models present stable equatorial circular orbits, as required by
consistency. Although the epicyclic approximation allows us to construct an approximate third integral near
a stable circular orbit
by adiabatic invariance of the corresponding vertical action \citep{bt}, the range of validity of this integral is very limited.
In particular, as we show below, the adiabatic approximation (AA) is valid for a region way much smaller than the thickness
of the Galactic thin disk. On the other hand, numerical experiments with orbits in axisymmetric disks have shown
throughout the years that the region of quasi-integrability around the stable circular orbit is much wider than
the region guaranteed by linear stability or by the KAM theorem
\citep{binney11, hunter05, sanders12}. Moreover, most methods to obtain approximate third integrals involve
extensive numerical modeling \citep{binney11, sanders12, bienayme12}. Thus, it is difficult to relate the corresponding results
to the physical parameters of the system.

We present in this manuscript a new approximate third integral of motion, valid for orbits which cross the Galactic thin disk
but do not belong to it. This integral describes the shape of nearly equatorial orbits in terms of the
dynamical surface density of the thin disk.
The approach is based on our recent results about a third integral of motion for nearly equatorial orbits in
razor-thin disks \citep{vieirathin} and is valid for any sufficiently flattened axisymmetric disklike configuration.
Corrections due to the
presence of additional structures are described and the results are compared with numerical simulations, showing
good agreement in regions near the vertical edge of the thin disk. In an era of growing attention devoted to Galactic
\citep{juric08, veltz08, famaey12, steinmetz12} and extragalactic \citep{bershady10, bershady10b} surveys,
simple expressions for third integrals of motion valid beyond the usual AA may be a crucial ingredient to a deeper
understanding of the dynamics underlying the kinematical data obtained.

\section{Dynamics of disk-crossing orbits}

For an axisymetric razor-thin disk, adiabatic invariance of the vertical approximate action $J_z$ next to a
stable circular orbit leads to the fact that the vertical amplitude $Z$ of the perturbed
test-particle orbit is given by \citep{vieirathin}
  \begin{equation}\label{invariant}
   Z(R)=Z(R_{o})\left[\frac{\Sigma(R_{o})}{\Sigma(R)}\right]^{1/3},
  \end{equation}
where $\Sigma$ is the surface density of the mass layer. Here $R_{o}$ and $R$ represent two values of the radial coordinate
(corresponding to usual cylindrical coordinates) along the orbit.
In order to test the validity of Eq.~(\ref{invariant}) for nearly equatorial orbits, we must specify what we mean by the function
$\Sigma (R)$ in the case of non-negligible thickness.
This function must have the following two properties: (i) It must reduce to the surface
density of the thin disk in the limit of zero thickness; (ii) it must represent a reasonable surface density profile for the
3D disk. We expect that, for highly
flattened disks (with a vertical dependence for $\rho$ approaching a Dirac delta), the behavior of nearly equatorial orbits
would be similar to the razor-thin case. In particular, Eq.~(\ref{invariant}) should still be valid in
this case for an appropriate function $\Sigma(R)$. Our goal in the next paragraphs is to describe the appropriate
``surface mass density function''
for disks with non-negligible thickness and to determine the range of validity of Eq.~(\ref{invariant}).

We consider the ``integrated dynamical surface density'' of the thin disk as
\citep{bershady10, bershady10b, bienayme12, efth2008, holmberg04}
  \begin{equation}\label{sdens}
   \Sigma_I(R) = \int_{-\zeta}^{\zeta}\rho_{tot}(R,z)dz,
  \end{equation}
where $\rho_{tot}$ is the ``dynamical'' density profile of the thin disk:
  \begin{equation}\label{rho}
   \rho_{tot} = \sum_i \rho_i.
  \end{equation}
The subscript $i$ corresponds to each gravitating component, such as thin disk, thick disk, bulge and (dark matter) halo.
Here, $\zeta$ is the thickness of the thin disk, which may vary with galactocentric radius. For sufficiently flattened disks we
can consider $\zeta$ as a constant, and in practice we may extend the corresponding integral of the thin-disk density
$\rho_{thin}$ to infinity.

The bulge and disk parts of (\ref{sdens}) are the surface densities obtained from photometric studies of
disk galaxies \citep{bt, vandedrkruit11}. They are related to the luminosity
profile  by assuming a constant mass-to-light ratio for the disk \citep{freeman, bosma, deblok}.
Furthermore, they reduce to the surface density of the corresponding razor-thin disk
when the vertical dependence of $\rho_{thin}$ on $z$ approaches a Dirac delta. In
this way, the above expression for $\Sigma(R)$ satisfies our requirements. Moreover,
recent studies were able to obtain estimates for the total
surface mass density (\ref{sdens}) in the solar position, including dark matter (see \citealp{holmberg04} and references therein).
The dynamical surface density of several spiral galaxies was also recently obtained via the ongoing DISKMASS survey
\citep{bershady10, bershady10b}. Besides that,
since the star's trajectory has no preference for the gravitational field of any specific component in (\ref{rho}),
we will adopt the complete expression (\ref{sdens}) further on.

Therefore, we expect that expressions (\ref{invariant}--\ref{sdens}) work well for orbits whose amplitudes are comparable to the
vertical thickness $\zeta$ of the thin disk (which may depend on R).
If we consider orbits inside the thin disk, a more reasonable approximate third integral
would be given by the usual adiabatic approximation (AA) \citep{bt}. If the orbit has a vertical amplitude much smaller than
$\zeta$, a non-negligible part of the disk's density distribution
(the portion with values of $z$ higher than the orbit's amplitude) will generate a gravitational
force which tends to repel the particle from the equatorial plane.
However, Eq.~(\ref{invariant}), together the integrated surface density (\ref{sdens}), considers that every mass element
of the disk generates a gravitational field which tends to bring the particle back to the equatorial plane. In this way,
eqs. (\ref{invariant}) and (\ref{sdens}) give us values of $Z$ which are smaller than the ones obtained from the AA. Therefore,
in regions very close to the equatorial plane, where the AA works well, eqs. (\ref{invariant}) and (\ref{sdens})
underestimate the orbit's vertical amplitude. The error obtained is usually small, though.
Furthermore, orbits with large vertical amplitudes are mostly affected by the bulge or halo gravitational fields,
depending on their mean galactocentric radius.

%\section{Orbits in galactic disks}

\begin{figure}[h]
   \includegraphics[scale=0.47]{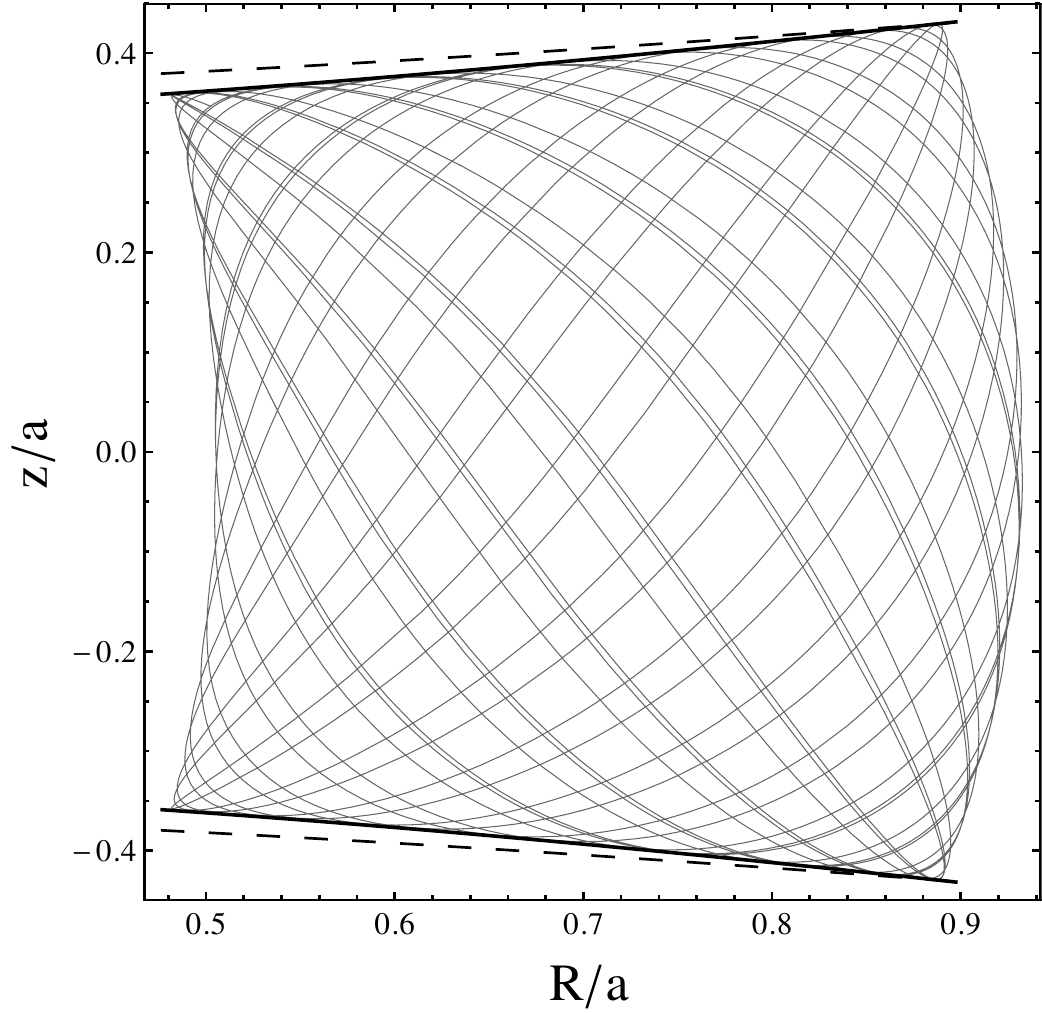}
   \caption{A typical quasi-circular orbit in the Miyamoto-Nagai potential with vertical amplitude close to
    $\zeta/a=3b/a$. The solid black line corresponds to the envelope predicted by (\ref{invariant}) with surface density
    (\ref{sdens}). The prediction from the adiabatic approximation (dashed line) overestimates the orbit's amplitude.
    We choose units in which $G=1$. $M=1$, $a=10$, $b=1$, $E=-0.055$, $L_z=0.8$,
    and the initial conditions are $R_0=5.05$, $z_0=1.1\times 10^{-1}$, $P_{R_0}=0$. The curves coincide at the point of maximum vertical amplitude, which is the starting reference to compute the predictions.}
    \label{fig:MNonecomp}
  \end{figure}

In fact, this is exactly what we obtain for a Miyamoto-Nagai potential (\citealp{MN}; see
Eq.~(\ref{helmimodel}) with $m=v=0$) and is illustrated in Fig.~\ref{fig:MNonecomp}.
These density distributions pervade all space. Nevertheless, we can define
a cutoff height by requiring that most of the disk mass is concentrated below this height. Although this last statement is not
precise, we see that for flattened disks (Eq.~(\ref{helmimodel})
with values of $b/a$ around $1/10$)
a good choice for the cutoff height is $\zeta\approx 3b$
%{\bf (and therefore $\zeta/a\approx 3b/a$ ) -- \%\%JAVIER, concorda? Mudei isso ao longo de todo o texto.\%\%}
, not depending on $R$.
The region between $-\zeta$ and $\zeta$ contains more than 95\% of the disk mass for any galactocentric radius
(the exact number varies with the ratio $b/a$).
It is interesting to remark that, for the Miyamoto-Nagai model,
Eq.~(\ref{invariant}) gives practically the same results when $\Sigma$ is the
surface density of the corresponding razor-thin disk ($b=0$) and when it is the
integrated surface density given by Eq.~(\ref{sdens}).
This occurs because the system is highly flattened ($b/a\approx 1/10$), in such a way that most of the
disk's density distribution is concentrated
in a very slim region near the equatorial plane.

\begin{figure}[h]
   \includegraphics[scale=0.47]{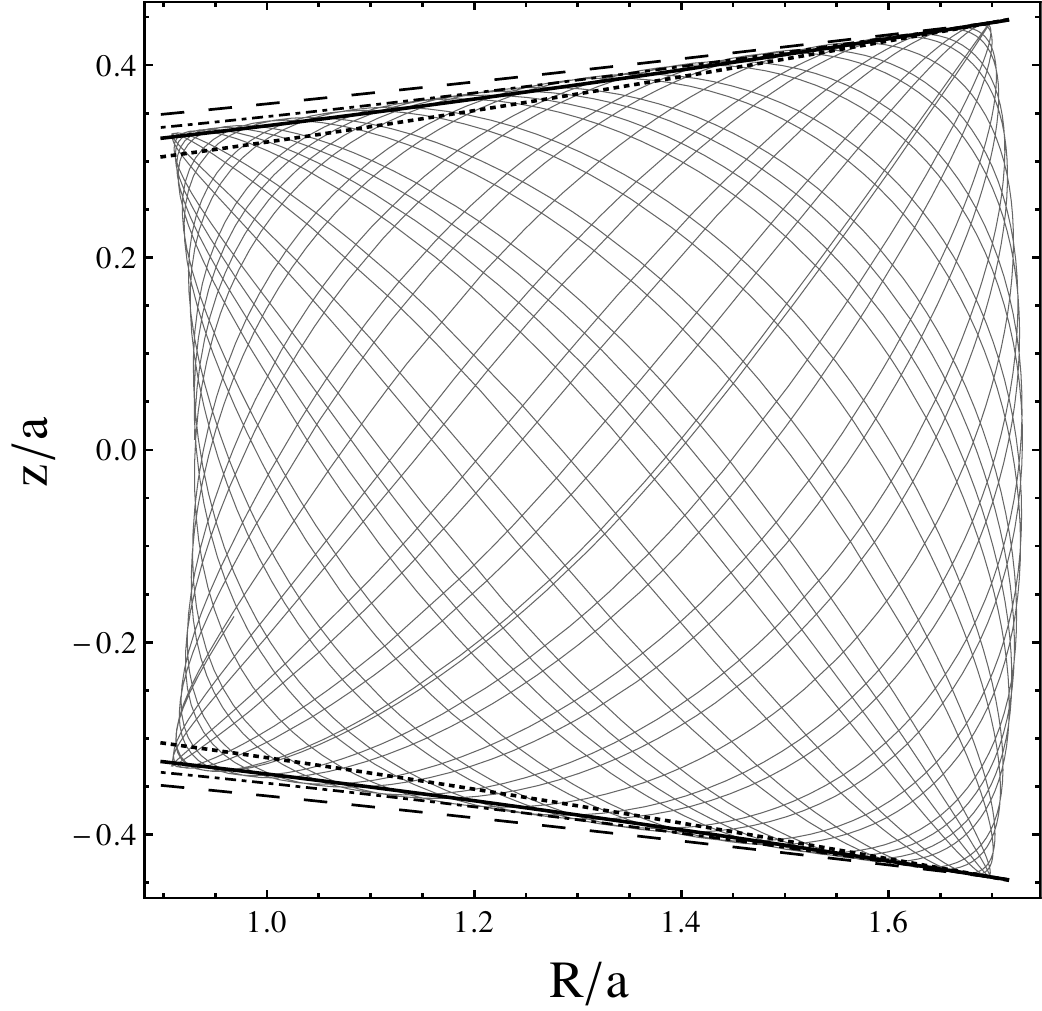}
   \caption{A typical quasi-circular orbit with large $L_z$
    in the two-component Miyamoto-Nagai model with vertical amplitude close to
    $\zeta/a=3b/a$. The solid black line corresponds to the envelope predicted by (\ref{invariant}) with surface density
    (\ref{sdens}), while the dashed line is the prediction from the adiabatic approximation.
    The dotted line corresponds to Eqs.~(\ref{invariant}--\ref{sdens}) with the contribution only of the thin disk,
    $\rho_{thin}$, and the dot-dashed line corresponds to the contribution only of $\rho_{thich}$.
    The thin-disk parameters are the same as in Fig.~\ref{fig:MNonecomp}.
    $E=-0.11$, $L_z=3$, $M_{thick}=2$, $a_{thick}=12$, $b_{thick}=4$,
    $R_0=9.3$, $z_0=1.1\times 10^{-1}$, $P_{R_0}=0$.}
    \label{fig:MNtwocomp}
  \end{figure}

If the thin-disk contribution dominates over the other components of the system, we can approximate
$\rho_{tot}\approx\rho_{thin}$. This approximation is very accurate if the density of the other components is very small or
absent, as tested in a two-component Miyamoto-Nagai model representing a superposition of a thin and a thick disk
(according to \citealp{hunter05}, these disks have a radial scale length $h=a+b$. Our simulations consider $h_{thin}<h_{thick}$).
However, as we increase the $z$- component $L_z$ of the orbit's specific angular momentum
-- therefore increasing the radius of the corresponding stable
equatorial circular orbit and, as a consequence, the mean radius of the 3D orbit -- corrections due to the presence of the thick
disk must be taken into account.
Furthermore, for higher values of mean galactocentric radius (where the thin-disk contribution
is very small compared to the corresponding thick-disk surface density), we can approximate Eq.~(\ref{sdens})
by the term due to $\rho_{thick}$.
According to the above numerical simulations, Eq.~(\ref{invariant}) still describes a very accurate
third integral of motion for amplitudes near $3b_{thin}/a_{thin}$.
We illustrate this behavior in Fig.~\ref{fig:MNtwocomp} with a typical quasi-circular
orbit of amplitude near the thin-disk cutoff, in the region where the thick-disk contribution begins to dominate.
The comparison bewteen different expressions for the approximate third integral shows us that Eqs.~(\ref{invariant}--\ref{rho}) result indeed in the most accurate description for the orbit's envelope.

In all cases, deviations from (\ref{invariant}) begin to appear once we consider orbits with higher energy. This fact is related
to another important remark: the above numerical experiments show that
our approximate integral does not work near resonances. In both situations, motion deviates significantly from the quasi-circular
approximation and higher-order terms in the potential become important.

Although our tests were performed for Miyamoto-Nagai disks, the above considerations lead us to the hypothesis that
this behavior is more general, working for any pair of axisymmetric thin+thick disk systems. In order to describe
the approximate third integral of motion (\ref{invariant}) in self-gravitating discoidal systems with more components
(such as a bulge and a halo) we must, in principle, include the contributions from each
different structure in (\ref{rho}).
These contributions must affect significantly the shape of nearly equatorial
orbits only in regions where their integrated surface density is comparable to the thin disk's surface density.

\section{A third integral of motion for disk stars in the Galaxy}

Recently, many models of the Galaxy appeared in the literature. The need to tackle some fundamental issues, such as the
degree of sphericity in the dark matter halo, led to simplifying hypotheses on the mass distributions of the different gravitating
components of the Milky Way. More specifically, some recent works described, for simplicity, the disklike part of the Galaxy
as a Miyamoto-Nagai profile \citep{allen91, johnston, helmi, irrgang13}. The potentials adopted for the halo
and for the bulge vary among authors. It is worthwhile to remark that
there is still some debate nowadays about the sphericity \citep{helmi, ibata13}
and triaxiality \citep{law09, ibata13, deg13} of the Galaxy's dark matter halo.

In this manuscript we consider	the ``bulge+disk+halo'' model described in \cite{helmi}. It consists of
 a Miyamoto-Nagai disk, a spherical Hernquist bulge \citep{hernquist90} and a
logarithmic potential for the halo. In view of the above discussion, we consider a spherical halo, case in which the
gravitational potential due to the Galaxy can be written as
\begin{equation}\label{helmimodel}
    \Phi=-\frac{G M}{\sqrt{R^{2}+\left(a+\sqrt{z^{2}+b^{2}}\right)^{2}}}
    -\frac{G m}{r+c}+v^{2}\ln\left(r^{2}+d^{2}\right),
\end{equation}
where $r^{2}=R^{2}+z^{2}$, $M=10^{11}M_{\bigodot}$ (disk mass), $a=6.5$ kpc and $b=0.26$ kpc are the disk parameters;
$m=3.4 \times10^{10}M_{\bigodot}$ (bulge mass), $c=0.7$ kpc are the bulge parameters; $v=131.5$ km s$^{-1}$,
$d=12$ kpc are the halo parameters.

Simulations in the corresponding ``disk+bulge'' system (i.e. by setting $v=0$) show us that the bulge
has a significant impact on the shape of disk-crossing orbits with amplitudes near $\zeta$ only in regions where its
contribution to (\ref{sdens}) is comparable to the disk's contribution. In these regions both expression
(\ref{invariant}) and the AA give poor predictions
for the shape of orbits, since the flattened component does not dominate anymore.
As the orbits get far from the bulge, the results described in the former paragraphs become valid again.
These results seem not to depend on the specific form of the bulge.
We also performed tests with a Plummer bulge \citep{bt} in place of the Hernquist profile and obtained the same qualitative scenario.

\begin{figure}[h]
   \includegraphics[scale=0.47]{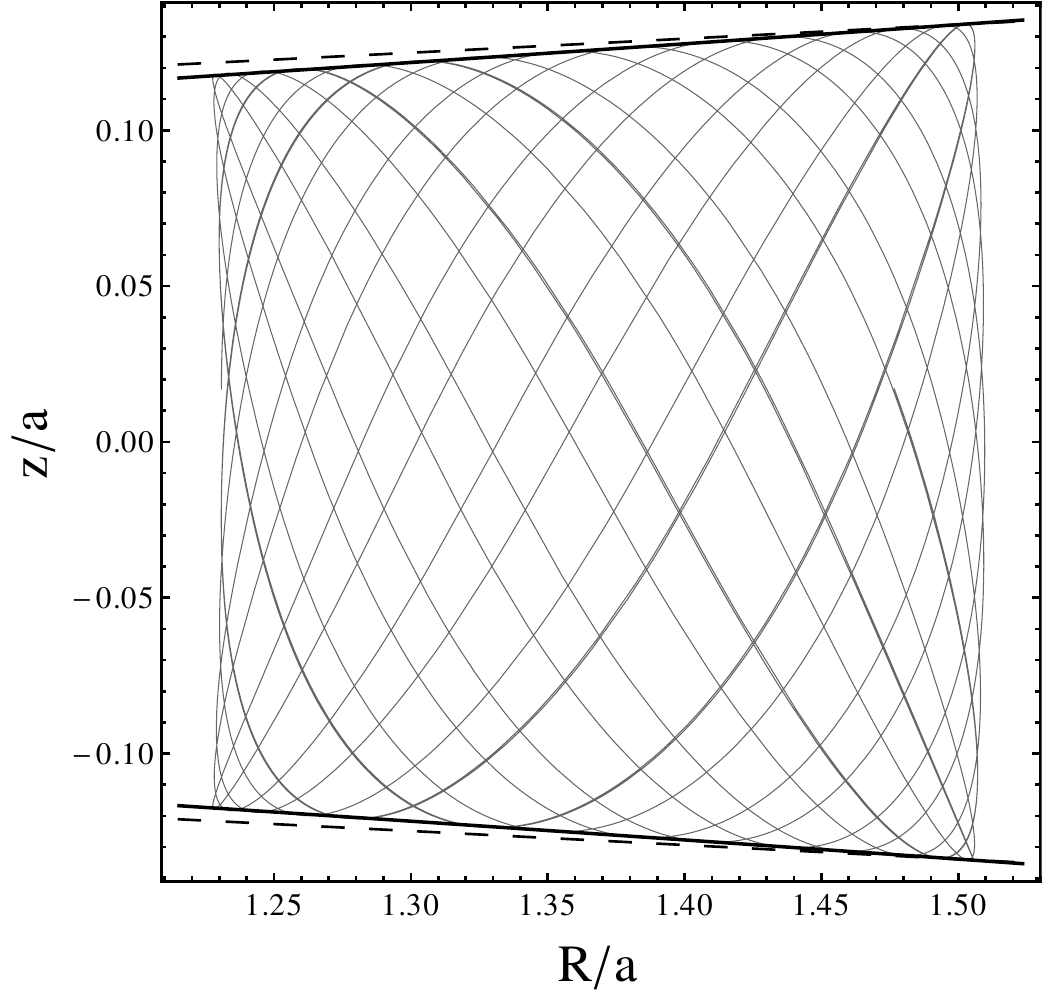}
   \caption{A typical quasi-circular orbit in Helmi's model \citep{helmi} with spherical halo.
    The orbit has $L_z=2000$  kpc (km/s), $E=65800$ (km/s)$^2$ ,
    $R_0=8$ kpc, $z_0=1.1\times 10^{-1}$ kpc, $P_{R_0}=0$,
    and lies near the solar neighborhood
    ($R\approx 8$ kpc, see \citealp{mcmillan11}), with vertical amplitude close to
    $\zeta/a=3b/a$. The solid black line corresponds to the envelope predicted by (\ref{invariant}) with surface density
    (\ref{sdens}), while the dashed line is the prediction from the adiabatic approximation.
    }
    \label{fig:helmi}
  \end{figure}

Numerical experiments show that the halo does
not affect significantly the prediction of (\ref{invariant}--\ref{sdens}) and our predictions are a good approximation for
orbits which do not get close to the bulge (see Fig.~\ref{fig:helmi}).
We also considered orbits in a Miyamoto-Nagai disk superposed by two Plummer spheres, which represent the bulge and the halo.
The results obtained are also in accordance with the general picture described above.
Energy conservation was checked for all calculations performed
with an accuracy characterized by a maximum relative error of $10^{-8}$.

 %===========================================================================
 %===========================================================================
 \subsection{Comparison with other approximations}

In this paragraph we examine whether there is a relation between our formula and the recently proposed correction to the adiabatic approximation
\citep{binney11}, where a term proportional to the vertical action, $\gamma J_z$, is added to the azimuthal angular momentum $|L_z|$ in order to
incorporate more accurately the centrifugal force contribution to off-equatorial orbits. The centrifugal term used to estimate the radial action
would depend, in that case, on $\mathcal{L}^2$ instead of $L_z^2$, where $\mathcal{L}=|L_z|+\gamma J_z$ and $\gamma$ is a constant.

As described in \citet{binney11}, there is in general a value of $\gamma$ for which the corresponding correction describes well the vertical amplitude of a
given off-equatorial orbit. This value, however, is found to be different for different orbits in the same potential, even in the case where the orbits
have the same value of energy and angular momentum (and with similar radial span).
In particular, it does not depend only on the flatness of the disk.
In our tests with a Miyamoto-Nagai potential ($b/a=0.1$)
we found that $\gamma$ should vary at least by one order of magnitude in order to correctly model orbits with different vertical
amplitudes, from $0.5\zeta$ to $1.5\zeta$
(still keeping $E$ and $L_z$ fixed for all orbits).

The factor $\gamma$ does not seem to present any clear correlation with the value of the vertical action,
although it seems to decrease as we consider orbits with higher vertical amplitude (however, orbits with the same amplitude
but with different $L_z$ are described by different values of $\gamma$).
For the orbit of Fig.~\ref{fig:MNonecomp}, for instance, one must have $\gamma\approx 6$ in order to describe accurately the orbit's envelope
(which coincides visually with the solid curve given by (\ref{invariant})). We must remark, however, that the above correction can give an accurate
results for orbits which are ``inside'' the disk (where the AA overestimates the amplitude and (\ref{invariant}) underestimates it)
and also orbits ``outside'' it (with amplitudes of roughly two times the disk thickness), once the corresponding value of $\gamma$ is found.

Although in this case vertical amplitudes can be described with accuracy,
the procedure of correcting the centrifugal contribution by considering also the orbit's vertical action (as in \citealp{binney11}) seems more intrincated
than prescribing one unique value of $\gamma$ for all orbits. As a conclusion, a more powerful prescription using this method would have to somehow
describe the dependence of $\gamma$ on the orbit's parameters, 
in such a way that {\it a priori} estimates of the orbit's vertical amplitudes could be made.

 %===========================================================================
 %===========================================================================

\subsection{Tests with completely integrable models}

In order to rigorously test the validity of (\ref{invariant}), we have to deal with 
potentials with an exactly conserved third integral of motion. An important and
 well known example is the case of St\"{a}ckel models, which have a simple separable form
in spheroidal coordinates (see \citealp{Batsleer-Dejonghe,bt}). For simplicity we will focus
on the Kuzmin-Kutuzov potential which can be written as
\begin{equation}\label{KuzminKutuzov}
    \Phi=-\frac{GM}{\sqrt{R^{2}+z^{2}+a^{2}+c^{2}+2\sqrt{a^{2}c^{2}+c^{2}R^{2}+a^{2}z^{2}}}},
\end{equation}
where $a$ and $c$ are real constants. As occurs in any St\"{a}ckel model, its orbits are bounded
vertically by  coordinate hyperbolae. In this case, they are defined by the equation
\begin{eqnarray}\label{hiperbola}
    2\lambda &=&a^2+c^2+R^2+z^2  \\ & &-\sqrt{z^4-2 \left(a^2-c^2-R^2\right) z^2+\left(a^2-c^2+R^2\right)^2},\nonumber
\end{eqnarray}
for each constant value of $\lambda$.
Moreover, the orbits are bounded radially by coordinate ellipses  given by
\begin{eqnarray}\label{elipse}
    2\nu &=& a^2+c^2+R^2+z^2  \\ & & +\sqrt{z^4-2 \left(a^2-c^2-R^2\right) z^2+\left(a^2-c^2+R^2\right)^2},\nonumber
\end{eqnarray}
for each constant value of $\nu$. Therefore, we have to compare the predictions of (\ref{invariant}) with the values
of vertical amplitudes determined by relation (\ref{hiperbola}). We will show that the choice $\zeta\rightarrow 0$, in
the integrated density of (\ref{sdens}), gives good numerical results.

\begin{figure*}[h]
 $$
 \begin{array}{cc}
 (a)$\qquad\qquad\qquad\qquad\qquad\qquad\qquad$ &\\
  \includegraphics[scale=0.42]{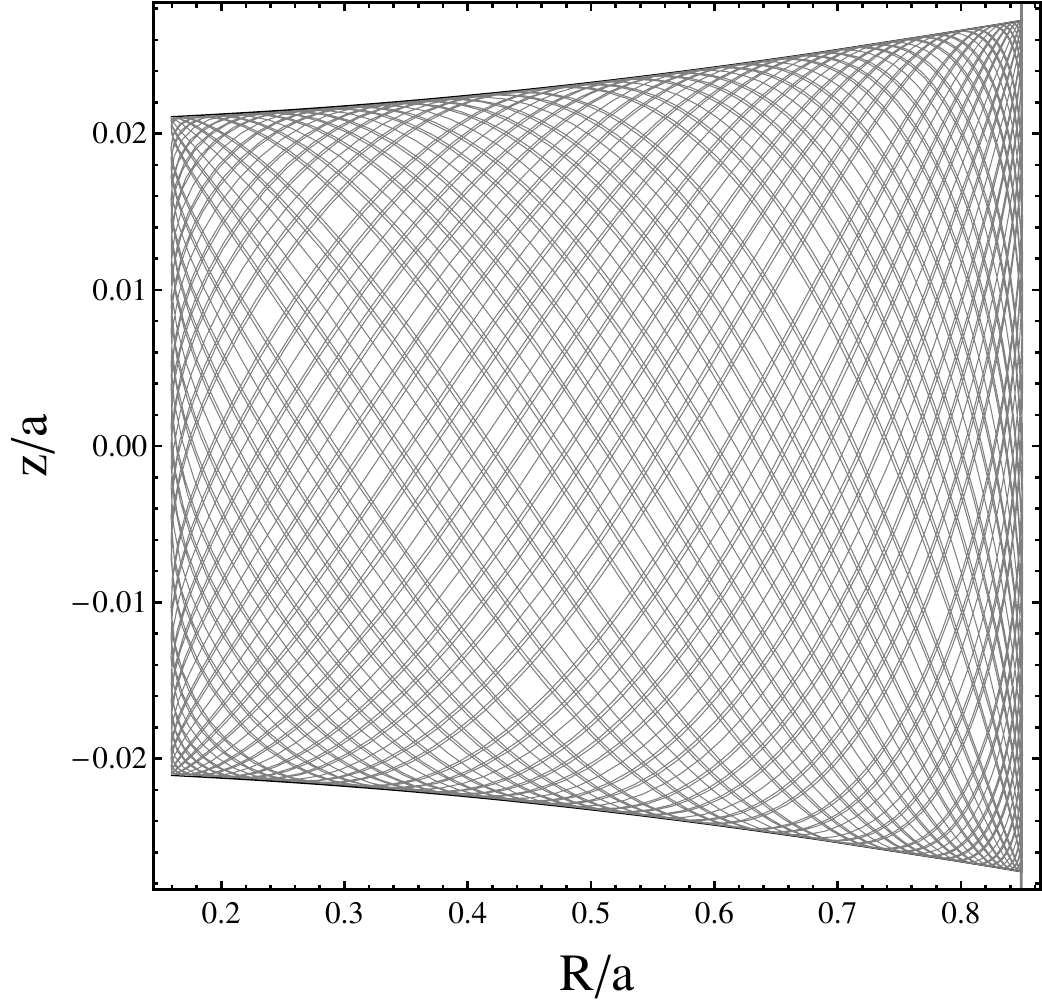} & \includegraphics[scale=0.42]{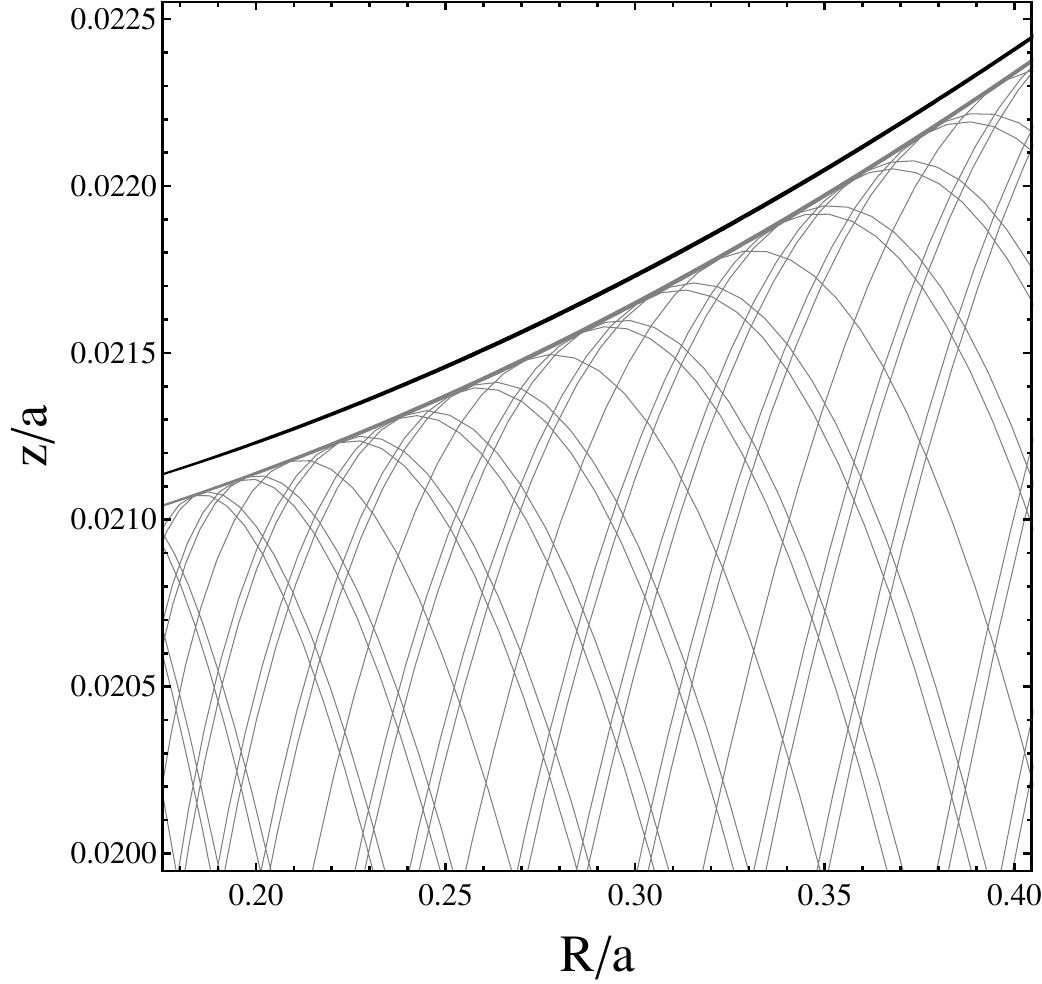} \\%0.5%
  (b)$\qquad\qquad\qquad\qquad\qquad\qquad\qquad$ &\\
   \includegraphics[scale=0.42]{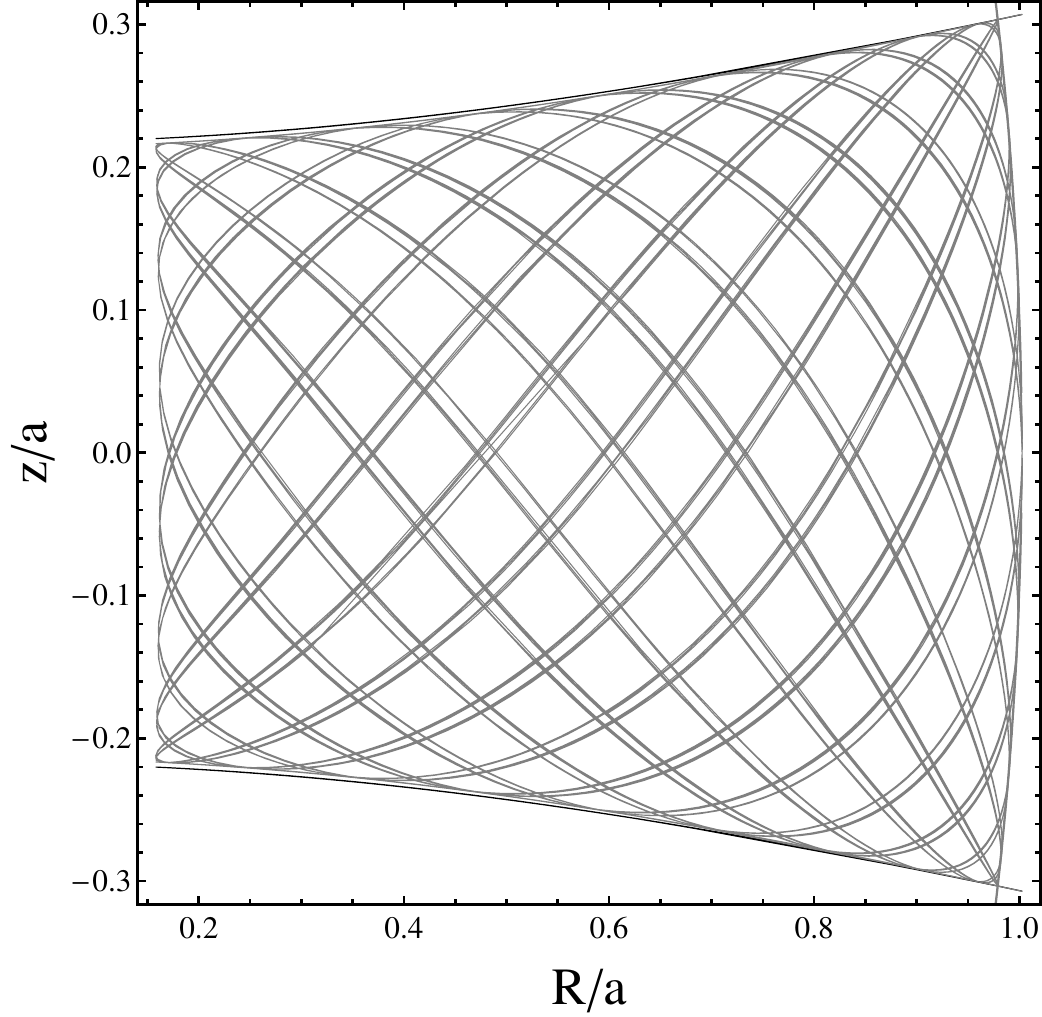} & \includegraphics[scale=0.42]{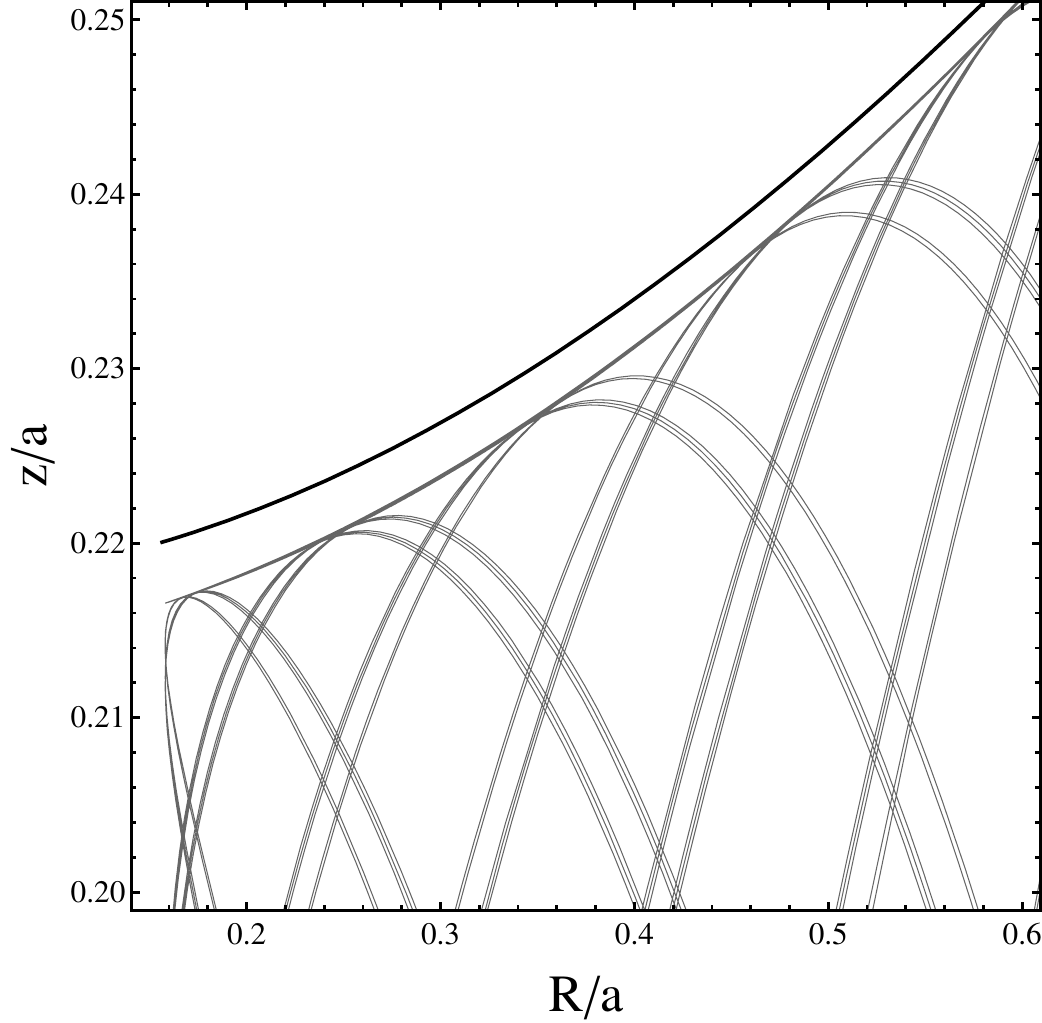}\\%2%
    (c)$\qquad\qquad\qquad\qquad\qquad\qquad\qquad$ &\\
  \includegraphics[scale=0.42]{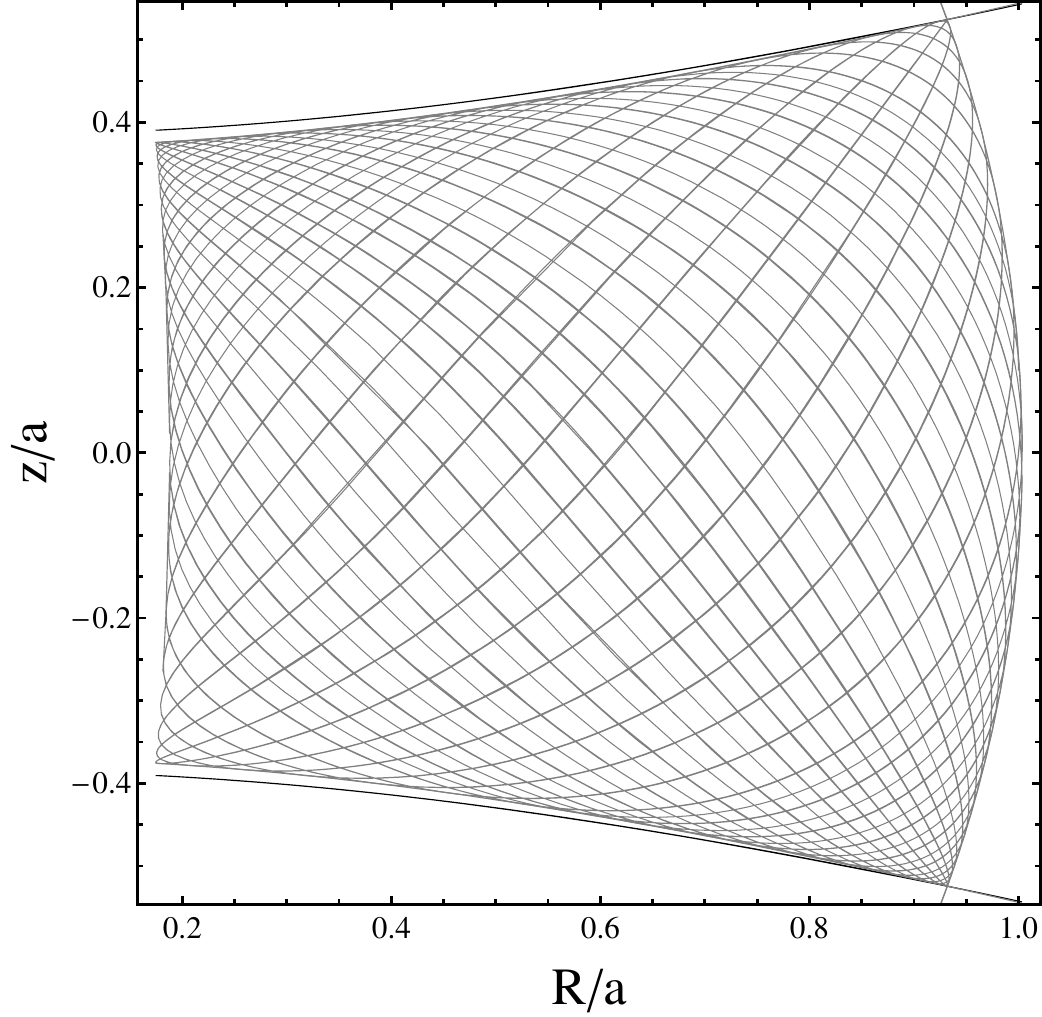} & \includegraphics[scale=0.42]{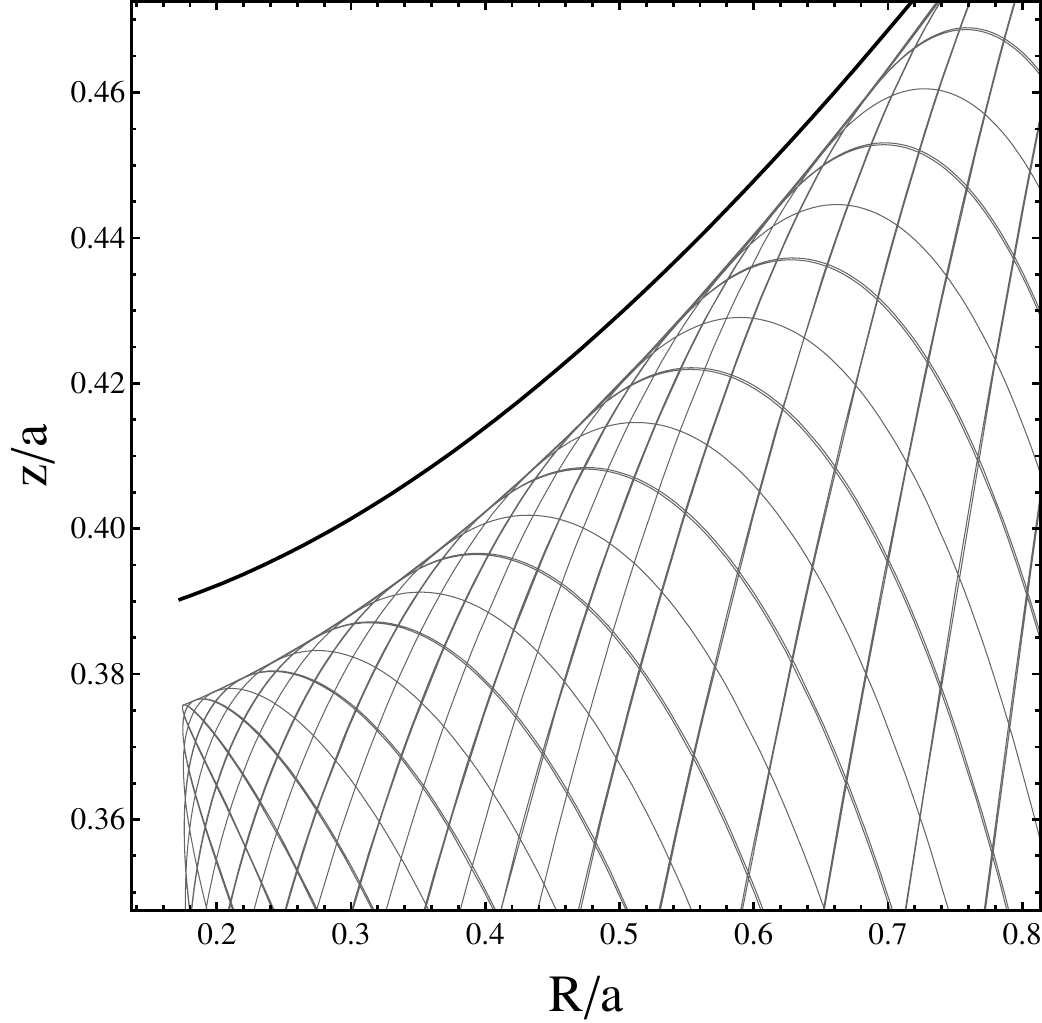}%4%
\end{array}
 $$  
   \caption{A sequence of three orbits with $L_{z}/\sqrt{GMa}=0.1$ in the Kuzmin-Kutuzov potential with $c/a=0.1$. Left panels 
   show the orbit in the meridional plane along with the envelope formed by the curves coordinates (gray) and
   the envelope predicted by relation (\ref{invariant}) (black). Right panels show a detail of the figures in the left, from
   which can be viewed that the difference between the grey and black envelopes increases with the vertical amplitude. The
   parameters for each orbit are:
   (a) $E a/GM=-0.7$,  $R(0)/a=0.85$, $z(0)/a=0.01$;
   (b) $E a/GM=-0.6$,  $R(0)/a=1.0$, $z(0)/a=0.1$; (c) $E a/GM=-0.55$,  $R(0)/a=1.0$, $z(0)/a=0.1$. All orbits have $P_{R_0}=0$.}
    \label{fig:KuzminKutuzov1}
  \end{figure*}
It is reasonable to expect  that (\ref{invariant}) works well for  vertical amplitudes of the order 
of  the disk thickness. In the model described by (\ref{KuzminKutuzov})
the thickness can be represented approximately  by the parameter $c$, so we can expect good predictions for
orbits with $Z/a\sim c/a$ or less. This fact is confirmed by numerical simulations as the corresponding to Fig.~\ref{fig:KuzminKutuzov1}, showing three typical cases of increasing vertical amplitude: (a) $Z/a\sim c/10a$, (b) $Z/a\sim 2c/a$
and (c) $Z/a\sim 4c/a$. We see that the difference between the envelope determined by (\ref{invariant}) and the one formed by
the coordinate hyperbolae increases with the vertical amplitude. The maximum percentage difference between envelopes black and gray
is approximately $0.5\%$, $2\%$ and $4\%$ for the cases (a), (b) and (c), respectively.

\begin{figure*}[h]
 $$
 \begin{array}{cc}
 (a)$\qquad\qquad\qquad\qquad\qquad\qquad\qquad$ &\\
  \includegraphics[scale=0.42]{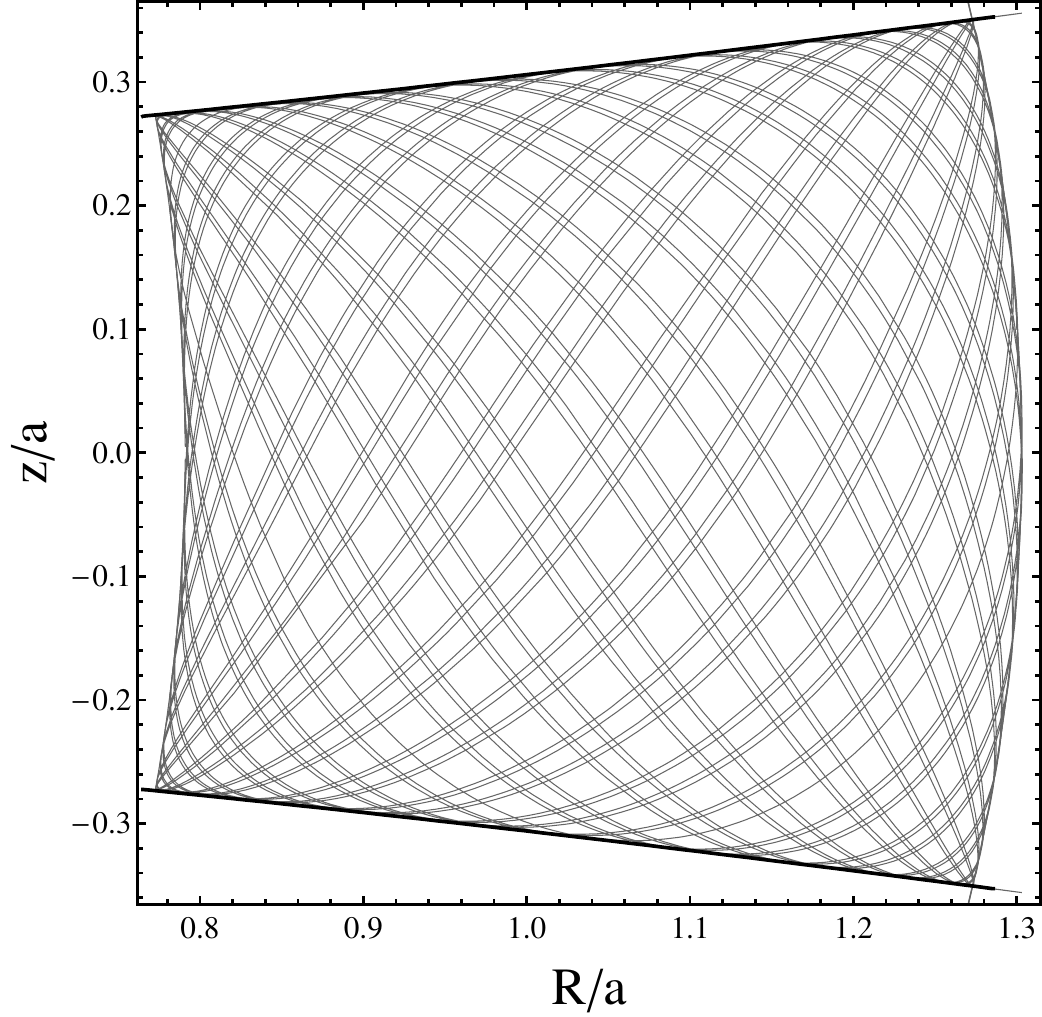} & \includegraphics[scale=0.42]{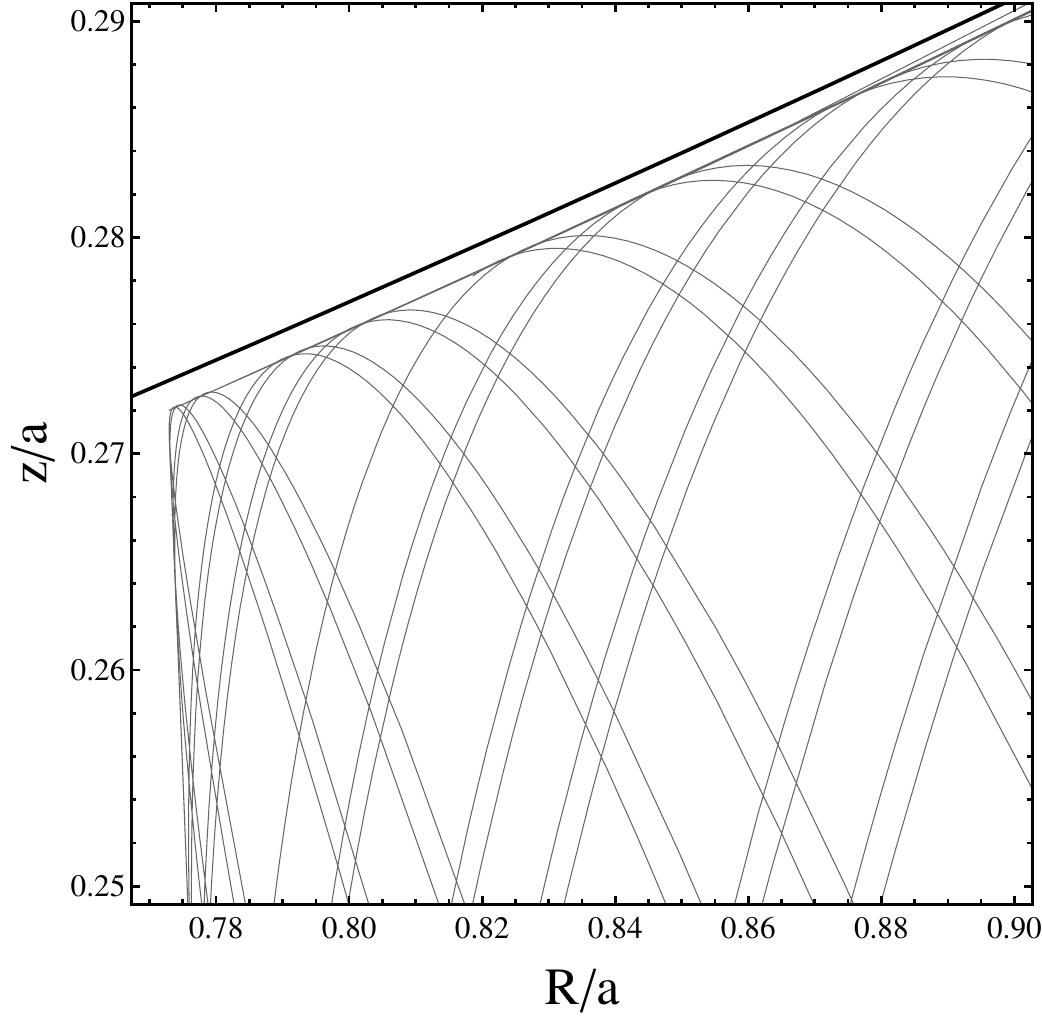} \\%0%
  (b)$\qquad\qquad\qquad\qquad\qquad\qquad\qquad$ &\\
   \includegraphics[scale=0.42]{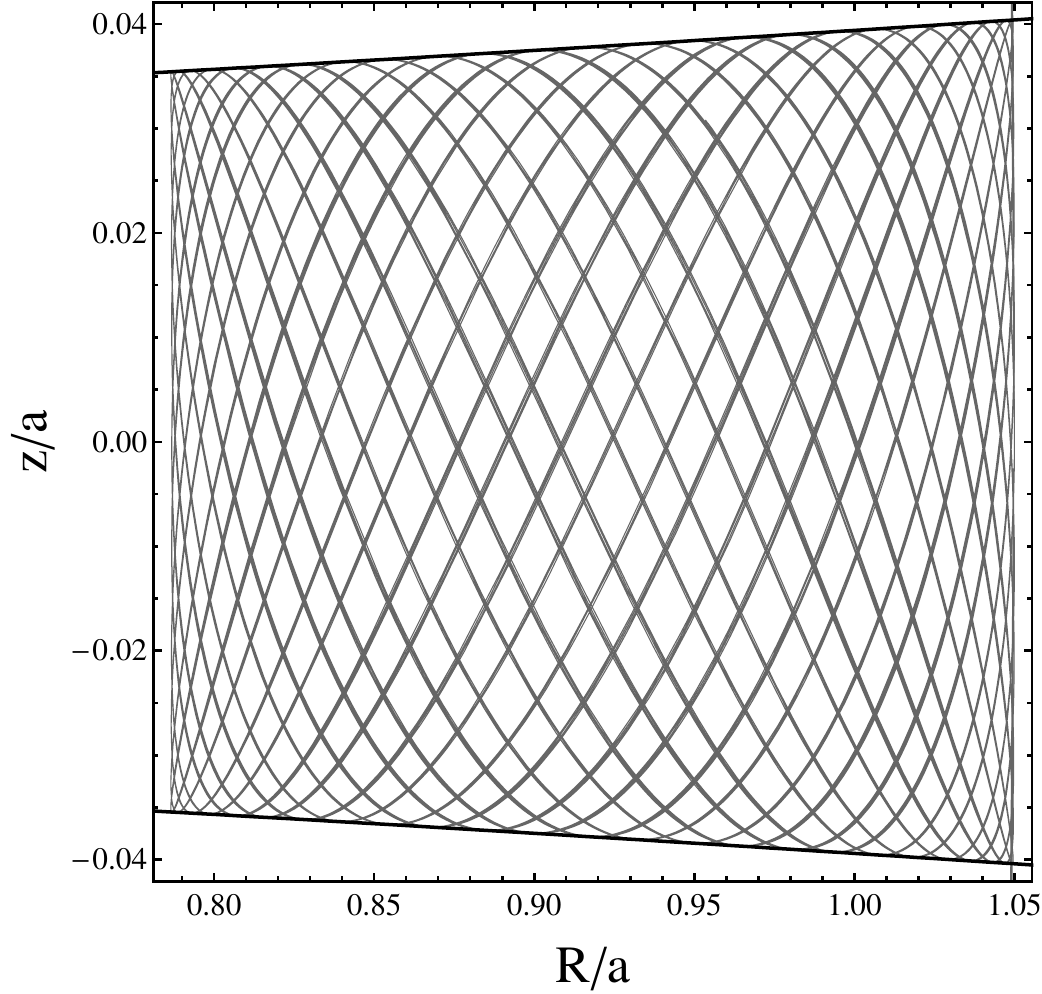} & \includegraphics[scale=0.42]{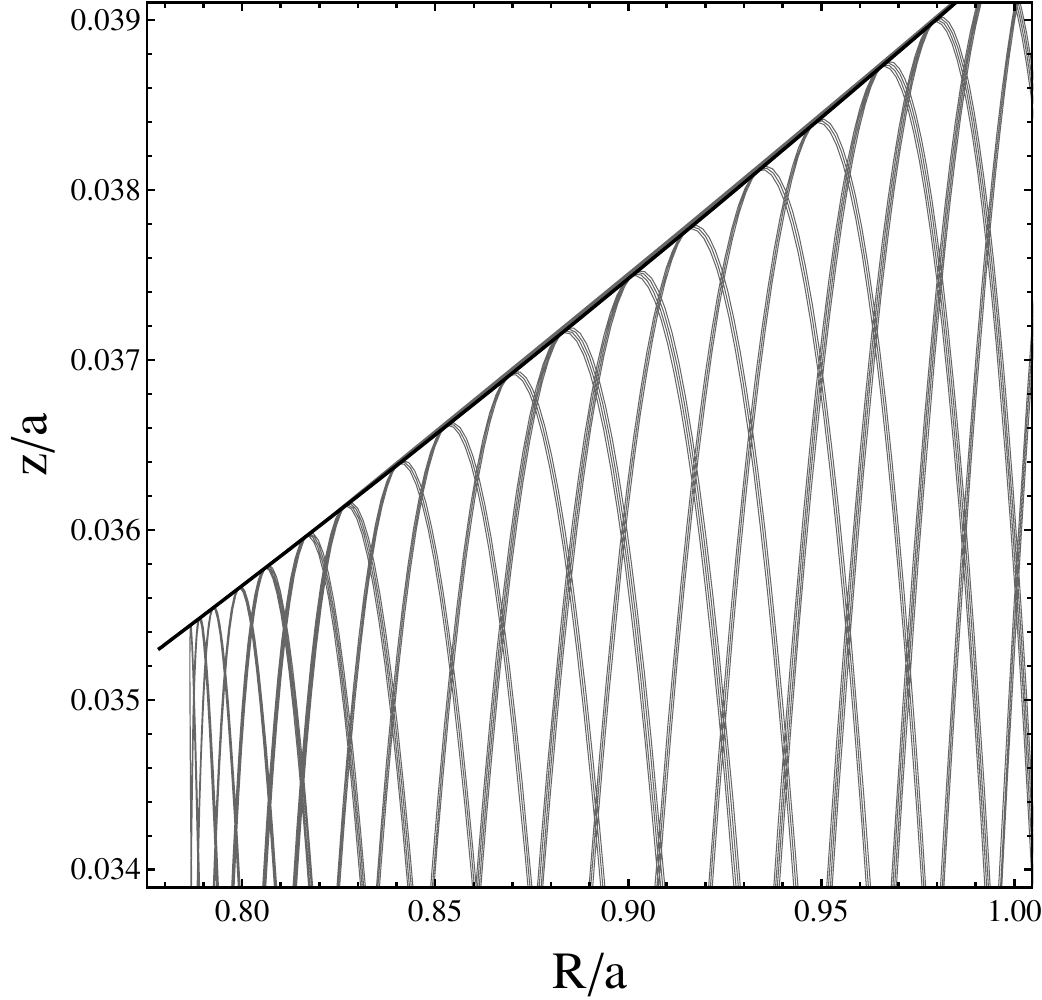}%0.3%
\end{array}
 $$
   \caption{Two orbits with $L_{z}/\sqrt{GMa}=0.5$ in the Kuzmin-Kutuzov potential with $c/a=0.05$. In case (a), where 
   $E a/GM=-0.46$,  $R(0)/a=1.3$ and $z(0)/a=0.1$, the envelope determined by (\ref{invariant}) (black) is very near the coordinate
   hyperbola (see the detail in the right panel), although the vertical amplitude is large compared to the thickness of the disk.
   A case with a smaller amplitude, of the order of $c/a$, as in case (b), reveals a very good prediction of (\ref{invariant}). In this case, $E a/GM=-0.55$,  $R(0)/a=1.05$, $z(0)/a=0.01$ and $P_{R_0}=0$.}
    \label{fig:KuzminKutuzov2}
  \end{figure*}

The predictions of (\ref{invariant}) improve for increasingly flattened disks. This can be seen, for example, by decreasing the
value of $c/a$ to $0.5$ (half of the value in Fig.~\ref{fig:KuzminKutuzov1}). 
Figure~\ref{fig:KuzminKutuzov2}a shows an orbit with a vertical amplitude ten times larger than
the thickness of the disk but with an envelope very near the coordinate hyperbola. The next orbit of Fig.~\ref{fig:KuzminKutuzov2}b,
with an amplitude of the order of $c/a$, exhibits an envelope coinciding completely with the coordinate hyperbola.
The maximum percentage difference between envelopes black and gray
in Fig.~\ref{fig:KuzminKutuzov2}a is approximately $0.3\%$, whereas for \ref{fig:KuzminKutuzov2}b there is no difference.
For smaller values of the ratio $c/a$ this behavior holds, tending to decrease the percentage difference in orbits with great amplitudes.

 %===========================================================================
 %===========================================================================

\section{Perspectives}

\subsection{Other Realistic mass models}

It would be instructive to check the validity of expressions (\ref{invariant}-\ref{sdens}) for orbits just outside the thin disk
in mass models with two-component stellar disks which take into account the exponential dependence of $\Sigma$ on
galactocentric radius \citep{mcmillan11}, as well as vertical profiles which are more consistent with recent
kinematic observations \citep{veltz08, juric08, steinmetz12}. For example, the luminosity density distribution within a galactic disk
can be written in the general form \citep{kruit1988,vandedrkruit11}
\begin{equation}\label{luminosity}
    L(R,z)=L_{o}e^{-R/h} \mbox{sech}^{2/n}\left(\frac{n z}{2 h_{z}}\right),
\end{equation}
where $L_{o}$ is the central luminosity density, $h$ is the scale length, $h_{z}$ is the scale height and $n=1,2,...$, i.e. ranging from the
isothermal distribution ($n=1$) to the exponential disk ($n\rightarrow\infty$).
By assuming that the mass-to-light ratio is constant, the
above mathematical expression can be used to represent also the stellar mass distribution, which is dominant in regions within the
optical disk and far from the bulge. In these regions,
%%%%%%%%%%%corresponding to the maximum disk,
the contribution of bulge and halo
can be neglected and $\Sigma_{I}$  will be proportional to $h_{z}e^{-R/h}$.
Observational results suggest that $h_{z}$ is independent of galactocentric radius \citep{vandedrkruit11} but, for the sake of generality, we
may consider it as a function of $R$. Thus,  equation (\ref{invariant})
reduces to
\begin{equation}\label{invariant-sec}
   \frac{Z^{3}(R)}{Z^{3}(R_{o})}=\frac{h_{z}(R_{o})}{h_{z}(R)}e^{(R-R_{o})/h}
\end{equation}
once we assume $\zeta(R)\propto h_z(R)$ in (\ref{sdens}). Eq.~(\ref{invariant-sec})
relates the vertical amplitude of disk-crossing orbits with the scale height (depending of $R$) and the scale length.
If $h_{z}$ is assumed to be constant, the prediction for the vertical amplitude is given by $Z(R)=Z(R_{o})e^{(R-R_{o})/3h}$.
It is worth pointing out that Eq.~(\ref{invariant-sec}) is valid for any value of $n$ in the luminosity profile (\ref{luminosity}).
In fact, Eq.~(\ref{invariant-sec}) is valid for any ``separable'' luminosity profile with a well-defined scale height $h_z(R)$.
If the radial profile is not strictly exponential, the exponential in Eq.~(\ref{invariant-sec}) should be substituted by
$L(R,0)/L(R_o,0)$. In regions where the thick-disk contribution becomes relevant, the corresponding third integral (\ref{invariant-sec})
will depend on $h_{z,thin}/h_{z,thick}$ -- as well as on $n$ -- in a nontrivial manner.

On the other hand, it is possible to relate the vertical amplitude of disk-crossing orbits with  the $z$-velocity dispersion
by combining (\ref{invariant}) with the equation of hydrostatic equilibrium \citep{vandedrkruit11},
\begin{equation}\label{hidro}
    \sigma_{z}(R)=\sqrt{c\pi G \Sigma_{I}(R)h_{z}},
\end{equation}
where $\sigma_{z}(R)$ is the $z$-velocity dispersion integrated over all $z$ and $c$ is some constant which varies between
$3/2$ (exponential disk, $n\rightarrow\infty$) to $2$ (isothermal distribution, $n=1$). If we again assume that $h_{z}$
is a function of $R$, then
\begin{equation}\label{invariant-disp}
   \frac{Z^{3}(R)}{Z^{3}(R_{o})}=\frac{h_{z}(R)}{h_{z}(R_{o})}\frac{\sigma_{z}^{2}(R_{o})}{\sigma_{z}^{2}(R)},
\end{equation}
which relates the vertical amplitudes of disk-crossing orbits to two mensurable quantities, $h_{z}$ and $\sigma_{z}$.
Note that for the case of galactic disks represented by the luminosity profile (\ref{luminosity})
we have that $\Sigma_I(R)\propto h_z(R)\exp(-R/h)$ and
the above expression reduces to (\ref{invariant-sec}).
In this case, $\sigma_z(R)/h_z(R)$ will have an exponential dependence on $R$
with an e-folding of twice the luminosity scale length, a result pointed out by
\citet{vandedrkruit11} in the case of constant scale height.

\subsection{Modified Theories of Gravity}

We now briefly describe the shape of orbits predicted by modified theories of gravity, as an extension of the
formalism presented in \citet{vieirathin}. If motion of test particles is described by a modified potential $\Psi$ (as in
\citealp{bekenstein1984, davi2010}) with
  \begin{equation}
   \frac{\partial\Psi}{\partial |z|}\bigg|_{z=0} = f \Sigma(R)
  \end{equation}
for a razor-thin disk, where $f(R,z)$ is a function depending on the parameters of the modified theory
and $\Sigma$ is the surface density of the thin disk,
then the extension of the razor-thin disk formalism to three-dimensional disks will generate an approximate third integral
of motion for nearly equatorial orbits which describes their envelope in the meridional plane by

  \begin{equation*}%\label{invariantmodified}
   Z(R)=Z(R_o)\left[\frac{\Sigma_\Psi(R_o)}{\Sigma_\Psi(R)}\right]^{1/3},
  \end{equation*}
with $\Sigma_\Psi$ defined by
  \begin{equation}\label{sigmapsi}
   \Sigma_\Psi(R) = \int_{-\zeta}^{\zeta}f(R,z)\rho(R,z)dz.
  \end{equation}
Here, $\rho(R,z)$ is the disk's density distribution. If $\rho$ is concentrated near the galactic equatorial plane and
if it falls off very rapidly with $z$, we can approximate $f$ in the integrand of (\ref{sigmapsi}) by its value at $z=0$.
With this approximation, the prediction for the envelopes in the meridional plane reduces to

  \begin{equation}\label{invariantmodified2}
   Z(R)=Z(R_o)\left[\frac{f(R_o,0) \Sigma_I(R_o)}{ f(R,0) \Sigma_I(R)}\right]^{1/3},
  \end{equation}
with $\Sigma_I$ given by Eq.~(\ref{sdens}). We note that Eq.~(\ref{invariantmodified2}) has the same form as its analogue
in the case of razor-thin disks \citep{vieirathin}, the only difference being the use of the integrated surface mass density
(\ref{sdens}) in place of the 2D surface density.

For MOND \citep{bekenstein1984}, Eq.~(\ref{invariantmodified2}) is given by (see \citealp{vieirathin})
  \begin{equation}\label{ZMOND}
   \frac{Z^{3}(R)}{Z^{3}(R_o)} = \frac{\mu_R}{\mu_{R_o}}\frac{\Sigma_I(R_o)}{\Sigma_I(R)},
  \end{equation}
where $\mu_R = \mu(|\nabla\Psi(R,0)|/a_o)$ is the MOND interpolating function, whereas for RGGR \citep{davi2010} we have
  \begin{equation}\label{ZRGGR}
   \frac{Z^{3}(R)}{Z^{3}(R_o)} =
\frac{1- V^2_\infty/\Phi_N(R_o,0)}
{1- V^2_\infty/\Phi_N(R,0)}\frac{\Sigma_I(R_o)}{\Sigma_I(R)},
  \end{equation}
where $\Phi_N$ is the Newtonian potential, given by Poisson's equation, and $V_\infty$ is the circular velocity at
infinity.

\section{Conclusions}

We extend the relation (\ref{invariant}), valid for axially symmetric razor-thin disks, to more general galactic models
including several components which usually are identified as bulge, thin disk, thick disk and halo. For this class of models
we can write
\begin{equation}\label{invariant2}
   Z(R)\Sigma_{I}^{1/3}(R)=Z(R_{o})\Sigma_{I}^{1/3}(R_{o}),
\end{equation}
where $\Sigma_{I}$ is the integrated dynamical surface density given by (\ref{sdens})--(\ref{rho}). Such a relation expresses
the fact that disk-crossing orbits are determined by an approximated third integral of motion of the form $I_{3}=Z\Sigma_{I}^{1/3}$.
By performing numerical calculations on a realistic model for the Milky Way (combinations of Miyamoto-Nagai models were also considered),
we show that the predictions of (\ref{invariant2}) are highly accurate for orbits far from the bulge and with vertical amplitudes
 of the order of the scale height. This fact suggests that (\ref{invariant2}) can be applied to the old stars belonging to the thick disk.

The above considerations may also be regarded as
a dynamical counterpart to distinguish between thin-disk and thick-disk stars in models of the Galaxy,
complementary to kinematic and
photometric studies (\citealp{pauli, veltz08, juric08}; see also \citealp{steinmetz12} and references therein).
Moreover, the approximate third integral obtained here can be used as an effective ``vertical action'' ($J_z=8 G^{(1/2)}/(3\pi^{1/2})\,I_3^{(3/2)}$; see \citealp{vieirathin}) in self-consistent
models for the Galaxy's distribution function depending on all action variables (e.g. \citealp{binney10, binney11}).

Finally, we have to remark that the predictions of Eqs.~(\ref{ZMOND}) and (\ref{ZRGGR}) may be used in the future as an additional test
of modified theories of gravity (MG), once sufficient data concerning the vertical amplitudes of orbits near the
equatorial plane of spiral galaxies becomes available. Even if predictions for the rotation curves are the same as in
Newtonian gravity + dark matter (DM) models, the behavior of off-equatorial orbits may be a
further probe to distinguish between DM and MG models for spiral galaxies.

\begin{acknowledgments}

R.S.S.V. thanks Funda\c{c}\~ao de Amparo \`a Pesquisa do Estado de S\~ao Paulo (FAPESP), grant 2010/00487-9, for financial support.

\end{acknowledgments}

\end{document}